\documentclass[twoside,a4paper,12pt]{article}
\usepackage{color}
\usepackage{amsfonts}
\usepackage[psamsfonts]{amssymb}
\usepackage[bottom]{footmisc}
\usepackage{amsmath}
\usepackage{multicol}
\newcounter{eg}                         \newtheorem{eg}{Example}[section]
\def\beg{\begin{eg}\rm}                 \def\eeg{\hfill\sq\end{eg}}

\usepackage{graphics}
\usepackage{graphicx}

\newcommand{\initiate}{\setcounter{equation}{0}} 
\def\Dirac{{\raise0.09em\hbox{/}}\kern-0.69em D}

\def\exterior{{{\raise0.2em\hbox{$\scriptstyle\bigwedge$}}{}}}

\def\kbar{{\mathchar'26\mkern-9muk}}

\def\lesssim{\mathrel{\hbox{\rlap{hbox{\lower8pt\hbox{$\sim$}}}\hbox{$<$}}}}

\def\sq{\hbox{\rlap{$\sqcap$}$\sqcup$}}

\def\p{\partial}

\def\dfrac #1#2{\displaystyle{\frac{#1}{#2}}}

\allowdisplaybreaks

\def\k{\kern-.1em\mathbin{,}\kern-.1em}
 
\def\hk{\kern.12em\raise-1em\hbox{$\hat{\raise1em\hbox{,}}$}\kern.12em}

\begin{document}

\title{Discrete fuzzy de~Sitter cosmology}

\author{Maja Buri\'c and Du\v sko Latas\thanks{majab@ipb.ac.rs, 
latas@ipb.ac.rs}
                   \\[15pt]
        University of Belgrade,  Faculty of Physics, P.O. Box 44
                   \\
        SR-11001 Belgrade
       }

\date{}
\maketitle
\parindent 0pt

\begin{abstract}
We analyze the spectrum of time observable 
in  noncommutative cosmological model introduced in
\cite{Buric:2015wta}, defined by $(\rho, s=\frac 12)\,$
representation of the de~Sitter group. 
We find that time has peculiar property: 
it is not self-adjoint, but  appropriate restrictions 
to the space of physical states give
self-adjoint extensions. Extensions
have discrete spectrum with logarithmic distribution
of eigenvalues, $\,t_n  \sim \ell\, \log\, n$+const, 
where  $\ell$ characterizes noncommutativity
and the usual assumption is $\,\ell=\ell_{Planck}$.
When calculated on physical states,  radius of the universe is
bounded below by $\, \ell\, \sqrt{\frac 34\, \left( \frac 14 
+\rho^2\right)}\, $, which resolves the big bang singularity.
An immediate consequence of the model 
is a specific  breaking of the original symmetry  at 
the Planck scale.
\end{abstract}

\setlength{\parskip}{10pt}

\pagestyle{plain}

\initiate
\section{Introduction}

The expression {`quantum space'}
 was introduced in the
early days of quantum mechanics by  Heisenberg, along
with  {`quantum derivative'} introduced by Dirac
who  observed that commutator is a derivation;
`points' of the quantum space are {`q-numbers'},  operators. 
Today the idea that spacetime, as seen by quantum particles,
is described by operators gives  strong heuristic 
and physical motivation for noncommutative geometry. 

There is a surprisingly simple covariantization of
the usual flat space of quantum mechanics 
 to  curved noncommutative spaces. If we
identify flat quantum space 
 with the Heisenberg algebra,
\begin{equation}
 [ip_i, x^j] = \partial_i x^j =\delta^j_i ,
\end{equation}
($\hbar =1$,  $p_i$, $x^j$ hermitian), curved quantum space
can be defined  by a moving frame  $e^\mu_\alpha\, $,
\begin{equation}
 [ip_\alpha, x^\mu] = e_\alpha x^\mu = e^\mu_\alpha(x)
                                   \label{tetrad}
\end{equation}
as in general relativity,  \cite{book}.
Adding to the last relation property (which one expects 
 in the quantum-gravity regime) that spacetime at the Planck 
scale is discrete or has a minimal 
quantum of  length, i.e. that coordinates may be non-commuting,
\begin{equation}
 [x^\mu, x^\nu] = i\kbar J^{\mu\nu}(x)\, ,
                                   \label{position}
\end{equation}
we have a general situation, a noncommutative algebra
of coordinates and momenta, $\cal{A}$. In principle, 
$\cal{A}\,$ may not have a Schr\"odinger-type representation 
of  momenta  through the partial derivatives; in fact, some
representations might be finite-dimensional. In this picture,
position algebra (\ref{position}) determines the 
structure  of the `points' of noncommutative space
i.e. the algebraic properties of coordinates, while (\ref{tetrad}) 
and the related commutators between momenta
define the differential-geometric structure and enable to 
introduce connection and curvature. Algebraic and geometric 
structures are intertwined  by the assumption that one
deals  with operators i.e. by associativity,  \cite{Buric:2006di}.

This is the general framework which we use.  
Its algebraic part is, in various  descriptions of noncommutative 
spaces,  more or less invariant, while the differential-geometric
part is specific in every approach: we use the noncommutative
frame formalism of Madore.
The frame formalism has proven in many aspects successful, in particular 
in describing spaces of euclidean signature with finite-dimensional 
representations like the fuzzy sphere and a number of other
 models in two and three dimensions, \cite{book,fs,Balachandran:2005ew}.
For further development of this concept it is crucial 
to provide realistic cosmological and astrophysical configurations
in four dimensions: this is the main motivation for our work.

Noncommutative geometry is but one of the approaches to
quantum gravity. Other approaches are perhaps, in the
view of description in terms of  lagrangian 
and quantization procedures, more fundamental. 
String theory introduces  elementary 
substructure which after quantization, macroscopically,
 gives spacetime geometry and classical gravity. In loop 
quantum  gravity,  vielbein and connection fields
 are basic variables which are quantized in 
background-space independent way. In these 
approaches `quantum space' with its properties 
is a derived quantity or notion. But in  most cases, being
effective or not,  coordinates are operators in the Hilbert
 space  of states: therefore  models, algebras  with
physically plausible features are  common to many theories.
We thus hope  that properties of
fuzzy de~Sitter space and its physical interpretation
discussed here will be of wider interest.

The plan of the paper is as follows. In Section~2
we introduce fuzzy de~Sitter space as a unitary irreducible
representation of the de~Sitter group, i.e. identify
its coordinates and differential structure.
In Section~3 we give  Hilbert space representation
for a specific  de~Sitter space defined by $(\rho, \frac 12)\,$
representation of the principal continuous series of $SO(1,4)$ and
solve the eigenvalue equation for the observable of 
cosmic time $\tau$. In Section~4 we examine the obtained solutions
and show how to redefine time to render it self-adjoint.
Finally, in the last section we discuss physical properties and
some cosmological implications of the given fuzzy geometry.

\section{Fuzzy de~Sitter space}

Our task is to study  observable of time 
in cosmological model introduced in \cite{Buric:2015wta, 
Buric:2017yes}. In
commutative geometry, four-dimensional de~Sitter space can 
be defined as an embedding in five-dimensional flat 
space \cite{Hawking},
\begin{equation}
 v^2-w^2-x^2-y^2-z^2 = -\mathsf{L}^2,        
\qquad
 ds^2=dv^2-dw^2-dx^2-dy^2-dz^2                    \label{embed}
\end{equation}
 where  $v\in(-\infty,\infty)$ is the embedding time.
Introducing
\begin{equation}
 \frac{\mathsf{t}}{\mathsf{L}} =\log\,\frac{v+w}{\mathsf{L}}\, ,\qquad
 \frac{\mathsf{x}}{\mathsf{L}} =\frac{x}{v+w}\, ,\qquad
 \frac{\mathsf{y}}{\mathsf{L}} =\frac{y}{v+w}\, ,\qquad
 \frac{\mathsf{z}}{\mathsf{L}} =\frac{z}{v+w}\,              \label{11}
\end{equation}
one obtains the line element in the FRW form,  the `steady 
state universe',
\begin{equation}
 ds^2 = d{\mathsf{t}}^2 -e^{\frac{2{\mathsf{ t}}}{\mathsf{L}}}\,
\left(d{\mathsf{ x}}^2+
d{\mathsf{ y}}^2+d{\mathsf{ z}}^2 \right)         .                       \label{22}
\end{equation}
Time $\,{\mathsf{t}}\in(-\infty, \infty)$ is defined only 
for $v+w>0$: coordinates (\ref{11}) cover only half of the 
de~Sitter space and the steady state space is incomplete, extendible.

Fuzzy de Sitter space can be defined in an analogous manner.
The general idea, realized in all details
for the fuzzy sphere \cite{fs}, is to identify spacetime
with the algebra of a Lie group, realizing the
embedding  through the Casimir relations: then in fact fuzzy 
spacetime is given by an irreducible representation of a Lie group.
We start with the group $ SO(1,4) $ with generators $M_{\alpha\beta}$
 ($\alpha, \beta = 0,1,2,3,4$),
\begin{equation}
 [M_{\alpha\beta}, M_{\gamma\delta}] = - i(\eta_{\alpha\gamma} M_{\beta\delta}
 -\eta_{\alpha\delta} M_{\beta\gamma}-\eta_{\beta\gamma}
 M_{\alpha\delta}+  \eta_{\beta\delta} M_{\alpha\gamma})   ,      \label{SO14}
\end{equation}
the signature is $\, \eta_{\alpha\beta}\,$= diag$(1,-1,-1,-1,-1)$.
Noncommutative extensions of $v$, $w$, $x$, $y$, $z\,$ are
embedding coordinates $\,x^\alpha$: they are proportional to the 
`Pauli-Lubanski vector' $W^\alpha$,
\begin{equation}
 W^\alpha =\dfrac 18 \,\epsilon^{\alpha\beta\gamma\delta\eta}
 M_{\beta\gamma} M_{\delta\eta} ,      \qquad
x^\alpha = \ell W^\alpha        \,  .                  \label{W}
\end{equation}
Dimensional constant   $\ell\,$ fixes the length scale of
noncommutativity: depending on physical interpretation, it 
can lie between the GUT scale and the Planck length
\cite{Buric:2017yes,Buric:2006nr}: usually one assumes
  $\,\ell \sim \ell_{Planck}$. One of the two Casimirs of $SO(1,4)$,
\begin{equation}
\eta_{\alpha\beta} W^\alpha W^\beta=
-{\cal W}
\end{equation}
 defines the embedding equivalent to (\ref{embed}). 
We will  for simplicity assume that the other Casimir operator
\begin{equation}
  {\cal Q} = -\frac 12 \, M_{\alpha\beta} M^{\alpha\beta}
\end{equation}
 is also fixed, i.e. that  fuzzy de~Sitter space is given by 
a unitary irreducible  representation (UIR) of the de~Sitter group.

All UIR's of the $ SO(1,4) $ are infinite-dimensional, 
labelled by two quantum numbers: conformal weight $\rho$ and spin $s$,
\cite{R},
\begin{equation}
 {\cal W} =  s(s+1)\big( \frac 14 + \rho^2\big),\qquad 
{\cal Q} = -s(s+1)+ \frac 94 + \rho^2 .              \label{Casimir}
\end{equation}
In the following we will use  UIR's of the principal continuous 
series,\,  $\rho\geq 0$\,,  $s = 0,\, 1/2,\, 1,\, 3/2\, , $
and the Hilbert space representations; in fact in this concrete
calculation we use only the simplest
nontrivial of them, $(\rho, s=\frac 12)$.

Various  choices of  differential calculi on  fuzzy 
de~Sitter space were discussed in \cite{Buric:2015wta}.
The simplest one which has the de~Sitter metric as commutative 
(macroscopic) limit is the calculus generated by four 
momenta, translations $\,ip_i= M_{i4}+M_{0i}\,$,
$\, i=1,2,3\, $ and dilatation $\, ip_0 =M_{04}\,$. When 
calculated,  expression (\ref{tetrad}) for  vielbein  
suggests to choose comoving coordinates 
 proportional to $\,W^i\,$,  and cosmic time
$\tau$  proportional to $\,\log(W^0+W^4)\,$ \cite{Buric:2015wta},
\begin{equation}
\frac{x^i}{\ell} = W^i  ,\qquad 
\frac\tau\ell = \log\,\frac{x^0+x^4}{\ell}=\log\, (W_0-W_4) \, .
\end{equation}

It is clear that  correct identification of  coordinates
 and momenta is very important for understanding of various 
properties and limits of a given fuzzy space, as well as
 for its physical interpretation. One way to see 
if noncommutativity improves the singularity structure
of  spacetime  is to determine the spectra of coordinates,
in this case $\tau$ and $x^i\,$, or $\sum (x^i)^2$.
As found in \cite{Buric:2017yes},  spectra
 of $\,x^i$ are continuous in $(\rho, s)\,$ 
representations;  embedding time $W^0/l\,$ 
has  discrete spectrum. Here we wish to
find  eigenvalues of the cosmic time.\footnote{An 
important observation is that  components $W^\alpha$
 are the Casimir operators of subgroups
of  $SO(1,4)$: $\,W^0$  of the $SO(4)$ and $\,W^i$
of the $SO(1,3)$.  Therefore, eigenvalues of  $\,W^\alpha$ could 
be in principle determined group-theoretically: by reduction
of a given UIR of the $SO(1,4)$  to the sum of UIR's 
of the corresponding subgroup. Similar strategy is
possible for  $\,\tau$ which is
one of two Casimir operators of the $E(3)$ subgroup, 
generated by  $\,  M_{0i}+ M_{i4}$ and $ M_{jk} $:
we have not succeeded to find the appropriate  
reduction formula in the literature.}

Properties of the spectrum  can be often  inferred 
 directly from the algebra. In this case we have relation
\begin{equation}
 [iM_{04}, W_0-W_4]= W_0-W_4 \, ,                          \label{MW}
\end{equation}
which implies that the group action of  dilatation $M_{04}$ is given by 
\begin{equation}
 e^{i\alpha M_{04}} \, (W_0-W_4) \, e^{-i\alpha M_{04}}=
e^\alpha(W_0-W_4).                                       \label{dilatation}
\end{equation}
The last formula means, apparently, that the spectrum of 
$\,W_0-W_4\,$ is continuous. Namely, it is easy to check 
formally that, if there is a nonzero  eigenvalue $\lambda>0$ 
of $\,W_0-W_4\,$ and the corresponding eigenvector
 $\,\psi_\lambda\, $,
\begin{equation}
 (W_0-W_4)\,\psi_\lambda= \lambda\,\psi_\lambda\, ,   \label{eigenv}
\end{equation}
then for every real $\alpha$,
$\, e^{-i\alpha M_{04}}\,\psi_\lambda\,$ is the eigenvector 
for the eigenvalue $\, e^\alpha\lambda\,$. This would mean that  
the spectrum consists of all real $\,\lambda>0$.
We will show in the following that eigenvalues of $\,W_0-W_4$,
 calculated in the Hilbert space representation
$(\rho, \frac 12\,)$, are in fact discrete. 
Namely,  differential equation (\ref{Eigen}) corresponding
to (\ref{eigenv}) has solutions of  finite norm
for all positive  $\,\lambda\in \mathbb{R}$, which, 
due to appropriate functional-analysis theorems, 
means that  $\,W_0-W_4\,$ is not self-adjoint.
The operator is only `formally symmetric'
because the domains of $\,W_0-W_4\,$ and $\,(W_0-W_4)^\dagger\,$ 
are not equal. There are, however, self-adjoint extensions
which we construct: each reduces the initial space of 
states to the `subspace of physical states', implying in 
consequence discreteness of  time.

\initiate

\section{Hilbert space representation}

We work in the Hilbert space representation of the
principal continuous series  $(\rho,s)$, \cite{Moylan}.
It is constructed in the familiar Bargmann-Wigner 
representation space of 
the Poincar\'e group with mass $m>0$ 
and spin $s$, \cite{Bargmann:1948ck}. Generators 
of the  Lorentz rotations are given by
\begin{equation}
 M_{\mu\nu} = L_{\mu\nu} + S_{\mu\nu}, \qquad \mu,\nu = 0,1,2,3\, ,
\end{equation}
 where $S_{\mu\nu}$ are spin generators, 
$\, L_{ik}= i\left( p_i\,\frac{\p}{\p p^k}-
p_k\,\frac{\p}{\p p^i}\right)\, $, 
$\ L_{0k}= i p_0\,\frac{\p}{\p p^k} $, $\,i, k=1,2,3\,$
and  $\,p_0=\sqrt{m^2+(p_i)^2}\,$.
Generators of the Poincar\' e translations, multiplication 
operators $\,p_\mu$, are used to define the remaining 
$\, M_{4\mu}\,$  by
\begin{equation}
 M_{4\mu} =\frac \rho m \, p_\mu -\frac{1}{2m}\, 
\left(p^\rho M_{\rho\mu}+M_{\rho\mu} p^\rho \right)  .
\end{equation}
This representation was used  in \cite{Buric:2017yes}: 
we will introduce it here very briefly in order 
to fix the notation and stress a couple 
of technical details and simplifications. 

Bargmann-Wigner space $\,{\cal H}\,$ for $s=\frac 12\, $ is  
the space of bispinors in momentum representation,
$\,\psi(\vec p)$, which are square-integrable solutions to 
the Dirac equation. Using  Dirac representation of 
$\gamma$-matrices, 
\ $ \gamma^0= \begin{pmatrix}
     I & 0\\ 0 & -I
 \end{pmatrix}$, \ $\gamma^i=  \begin{pmatrix}
     0 & \sigma_i\\ -\sigma_i & 0
 \end{pmatrix} $, 
 $\, \psi(\vec p)\,$ can be written as
\begin{equation}
 \psi(\vec p) =\begin{pmatrix}
     \Phi(\vec p)\\[4pt]
 -\,\dfrac{\vec p\cdot\vec \sigma}{p_0+m}\,\Phi(\vec p)
 \end{pmatrix}  
\end{equation}
where $\Phi(\vec p)$ is an unconstrained spinor.
Scalar product is given by
\begin{equation}
 (\psi,\psi^\prime)=
\int \frac{d^3p}{p_0}\, \psi^\dagger\gamma^0\psi^\prime
=\int \frac{d^3p}{ p_0}\,
 \frac {2m}{p_0+m}\, \, \Phi^\dagger \Phi^\prime  \, .
\end{equation}
Written in blocks of 2$\times$2 matrices, 
$M_{\alpha\beta}$ and $W_\alpha$  have the form
$ \  M=\begin{pmatrix}
     A & B\\
     B & A
 \end{pmatrix}\, $.
Matrix elements of such operators are\footnote{At
this point we fix the relative positions of $\gamma^0$,  $1/p_0$
and $M$: this ordering is not essential and can be changed, 
but implies appropriate changes in relations which follow.}
\begin{equation}
 (\psi, M\psi^\prime) = \int d^3p\, \,\psi^\dagger \,\frac{\gamma^0}{p_0}\,
M \psi^\prime = \int \frac{d^3p}{p_0}\,\, \Phi^\dagger \left( 
A-\frac{p_k\sigma^k}{p_0+m}\, A \, \frac{p_i\sigma^i}{p_0+m}
+[B, \frac{p_k\sigma^k}{p_0+m}]\right) \Phi^\prime . \nonumber
\end{equation}
Eigenvalue problem $\,  M\psi =\lambda\psi\, $
can be written as a set of two spinor equations:
\begin{eqnarray}
 &&  \left( 
A-\frac{p_k\sigma^k}{p_0+m}\, A \, \frac{p_i\sigma^i}{p_0+m}
+[B, \frac{p_k\sigma^k}{p_0+m}]\right)\Phi = \lambda\,  
\, \frac{2m}{p_0+m}\, \Phi  \,                          \label{zvezda}
\\[6pt]
&& \left( [A, \frac{p_k\sigma^k}{p_0+m}]
+B-\frac{p_k\sigma^k}{p_0+m}\, B \, \frac{p_i\sigma^i}{p_0+m}
\right)\Phi =0\, .
\end{eqnarray}
One can easily check that the second equation
is fulfilled for all solutions of the first, so essentially 
one has to solve (\ref{zvezda}).

In our  problem $\, M=W_0-W_4\,$, the blocks $A$ and $B$ are 
\begin{eqnarray}
 && A = -\,\frac{1}{2m} \left(\rho - \frac i2\right) p_i\sigma^i-\frac{i}{2m} \, 
p_0(p_0+m)\,\frac{\p}{\p p_i}\,\sigma_i\, ,\\[6pt]
&&B = -\, \frac {1}{2m}\epsilon^{ijk}(p_0+m)p_i\,\frac{\p}{\p p^j}\,\sigma_k -\, \frac{3i}{4m}\,(p_0+m)\, .
\end{eqnarray}
Eigenvalue equation for $\, W_0-W_4\,$ becomes
\begin{equation}
 \left(- \frac{\rho}{2m}\, p_i\sigma^i - \frac i2\,
(p_0+m)\, \frac{\p}{\p p_i}\, \sigma_i
 +\frac{i}{2m}\, p_i\, \frac{\p}{\p p_i}\, p_j\sigma^j \right)
 \Phi =\lambda\Phi  \,   .                           \label{Eigen}
\end{equation}
As $\, W_0-W_4\,$ commutes with  3-rotations $M_{ij}$, 
we can choose the eigenfunctions in the form
\begin{equation}
\Phi_{\lambda jm}(\vec p) = \frac{f(p)}{p}\,
\phi_{jm}(\theta,\varphi)+\frac{h(p)}{p}\,
\chi_{jm}(\theta,\varphi) ,
\label{aansatz}
\end{equation}
where $\,p\,$ is the radial  momentum, $\, p^2 =-p_i p^i= p_0^2 -m^2\, $,
and $\, \phi_{jm} $ and  $ \chi_{jm} $ are the
eigenfunctions of the angular momentum.
Using (\ref{aansatz}) we  obtain radial equations for $f$ and $h$: 
\begin{align}
& (p_0+1)\, \frac{df}{dp_0}+ i\rho f-\frac{ j+\frac 12 }{p_0-1}\,
  f =2i\lambda\, \frac hp \, ,                   \\[6pt]
&  (p_0+1)\, \frac{dh}{dp_0}+ i\rho h+\frac{ j+\frac 12 }{p_0-1}\,
  h =2i\lambda\, \frac fp   
\, . 
\end{align}

Solutions to these equations  are derived in Appendix 1. They are
expressed in terms of the Bessel functions using variable 
$ \,  z=\sqrt{\frac{p_0-m}{p_0+m}}\, $; this variable
varies in a finite interval, $\,  z\in(0,1)\,$.
Of two linearly independent solutions for fixed $\lambda$
and $j$  one is regular,
\begin{equation}
 f_{\lambda j} = C\left(\frac{2}{1-z^2}\right)^{-i\rho} \sqrt{z}\, 
J_j(2\lambda z),\quad\ 
h_{\lambda j} = iC\left(\frac{2}{1-z^2}\right)^{-i\rho} \sqrt{z}\, 
J_{j+1}(2\lambda z)            \, ,               \label{resenje}
\end{equation}
and therefore we conclude that the spectrum of $\, W_0-W_4\, $ is
 positive real axis, $\lambda\in (0,\infty)$.
However, the given set of solutions is not 
orthonormal. The scalar product of  two eigenfunctions is
\begin{eqnarray}
 (\psi_{\lambda jm},\psi_{\lambda' j'm'}) 
 =2\delta_{jj'}\delta_{mm'}C^*C'
\int\limits_0^1z dz\, \left( J_j(2\lambda z)J_j(2\lambda'z)+
 J_{j+1}(2\lambda z)J_{j+1}(2\lambda'z) \right) \, .       \label{Proizvod}
\end{eqnarray}
As Bessel functions  $\,J_j(\zeta)\,$ are finite in any finite interval,
 integral (\ref{Proizvod})  is bounded
for $\lambda =\lambda'\,$, i.e. all solutions are normalizable,
which is in contradiction with the statement that they belong
to continuous spectrum. 
Also they are not orthogonal for $\lambda\neq \lambda'$.
 Therefore, not all of  formal solutions  (\ref{resenje})
can be the eigenfunctions of a self-adjoint operator, and self-adjointness
is a property we would certainly like $\tau$ to have.

\initiate

\section{Self-adjoint extensions}

The obtained result 
requires  additional analysis. We  started with a
 unitary representation of the $SO(1,4)$, that is, with a set of 
self-adjoint   (hermitian) 
generators $M_{\alpha\beta}$. We defined $W_\alpha$
by (\ref{W}), as a sum of products of operators which mutually commute. 
Therefore formally, $\,W_0-W_4 =\tau/\ell\,\, $ is hermitian and 
should have an orthonormal eigenbasis (discrete or continuous). But
in concrete representation 
 we obtained a 
continuous set of eigenfunctions of finite norm which 
are not mutually orthogonal.  Hence $\tau$ is
not self-adjoint: it can only be {formally symmetric},
with domain ${\cal D}(\tau)$ unequal to the domain
of its adjoint, ${\cal D}(\tau^\dagger)$. To define 
self-adjoint extensions, if they exist, we need to 
resolve the  issue of the domains.

Problem is obviously in the radial equation.
Separation of angular variables gives a division of  $\,{\cal H}\,$ 
into subspaces of fixed angular momentum $j$, in which 
$\,\tau$  reduces to operators  $T_{j}$:
\begin{eqnarray}
&& \left(\psi_{jm},(W_0-W_4)\,\psi'_{j'm'}\right)\equiv
\delta_{jj'}\delta_{mm'}
\int\limits_0^1 dz\,\Phi^\dagger\, T_{j}\, \Phi'
\label{radial}                         \\
&& \  =   
2\delta_{jj'}\delta_{mm'}
\int\limits_0^1 dz\, 
\begin{pmatrix}
  f^* & h^*
\end{pmatrix}
\begin{pmatrix}
 0 &\rho\,\frac{2z}{1-z^2}-i(j+\frac 12)\frac 1z -i\, \frac{d}{dz} \\[2pt]
\rho\,\frac{2z}{1-z^2}+i(j+\frac 12)\frac 1z  -i\, \frac{d}{dz} & 0
\end{pmatrix}
\begin{pmatrix}
 f' \\ h'
\end{pmatrix}              \, \nonumber   \\[2pt]
&&\  = 
\delta_{jj'}\delta_{mm'}
\int\limits_0^1 dz\, 
\begin{pmatrix}
  F^* & H^*
\end{pmatrix}
\begin{pmatrix}
 0 & -i\, \frac{d}{dz} \\[2pt]
 -i\, \frac{d}{dz} & 0
\end{pmatrix}
\begin{pmatrix}
 F' \\ H'
\end{pmatrix}              \, \nonumber .
\end{eqnarray}

Functions $F$ and $H$ are defined by
\begin{equation}
 F=\left(\frac{2}{1-z^2}\right)^{i\rho}  z^{-j-\frac 12}\, f\, ,\qquad
 H=\left(\frac{2}{1-z^2}\right)^{i\rho}  z^{j+\frac 12}\, h\, ,
                                     \label{Newfunctions}
\end{equation}
and they are introduced in  Appendix 1 to solve the radial equation;
they simplify the matrix elements of $\,T_j$ as well as the
 scalar product,
\begin{eqnarray}
&& \left(\psi_{jm},\psi'_{j'm'}\right)=
2\delta_{jj'}\delta_{mm'}
\int\limits_0^1 dz\, 
\begin{pmatrix}
  F^* & H^*
\end{pmatrix}
\begin{pmatrix}
 z^{2j+1} & 0\\
  0 &z^{-2j-1}
\end{pmatrix}
\begin{pmatrix}
 F' \\ H'
\end{pmatrix}              \, \nonumber  .
\end{eqnarray}

Let us examine properties of $T_j$.
In order to find  $ T_{j}^\dagger$ we partially
 integrate (\ref{radial}),
 \begin{eqnarray}
&& \left( \psi_{jm},(W_0-W_4)\,\psi'_{j'm'}\right)
=-i\, \delta_{jj'}\delta_{mm'}
\int\limits_0^1 dz \left( F^*\frac{dH'}{dz}+H^* \frac{dF'}{dz} \right)
 \nonumber \\[0pt]
&& \phantom{\psi_{jm} \  }=
i\,\delta_{jj'}\delta_{mm'}
  \int\limits_0^1 dz\, \left( \frac{dF^*}{dz} H'+\frac{dH^*}{dz} F'\right)
-i\delta_{jj'}\delta_{mm'} \big( F^*H'+H^*F' \big) \left\vert_0^1 \right.
    \,.                                         \label{bound}
\end{eqnarray}
We see that  
the `action' of $T_j\, $ on functions, given by the first term
in (\ref{bound}), is self-adjoint: but since the boundary term 
does not  vanish,  $T_j$ and $T_j^\dagger$ are not equal. 
This is in fact a definition of being  `formally symmetric', 
\cite{pim}.  The other signature of 
non-hermiticity  are nonzero deficiency indices, i.e. the
existence of normalizable solutions to equations 
$ \ T_{j}\Phi =\pm i\Phi\,$.
We show in Appendix 2 that the deficiency indices of $\,T_j\,$ 
are $\,(n_+,n_-)=(1,1)$.

There is a systematic way of extending formally symmetric 
operators to the self-adjoint, \cite{pim,dunford}.
 The main idea is to find an appropriate subspace of $\cal{H}$
on which the boundary term  vanishes: this subspace becomes
the domain of both, redefined or `extended' $\tau\,$
 and $\tau^\dagger$.
A necessary condition for  existence of  self-adjoint extensions
 is that deficiency indices $n_+$ and $n_-$ be equal. Analyzing
 (\ref{bound}) in Appendix 2 we find that $T_j$
is self-adjoint if it is restricted to subspace of functions 
(\ref{Newfunctions})  which satisfy
\begin{equation}
F(0)=H(0)=0, \qquad 
  H(1)  = ic F(1)\,  ,                                \label{Key}
 \end{equation}

\begin{figure}[h]
\begin{center}
 \includegraphics[scale=0.5]{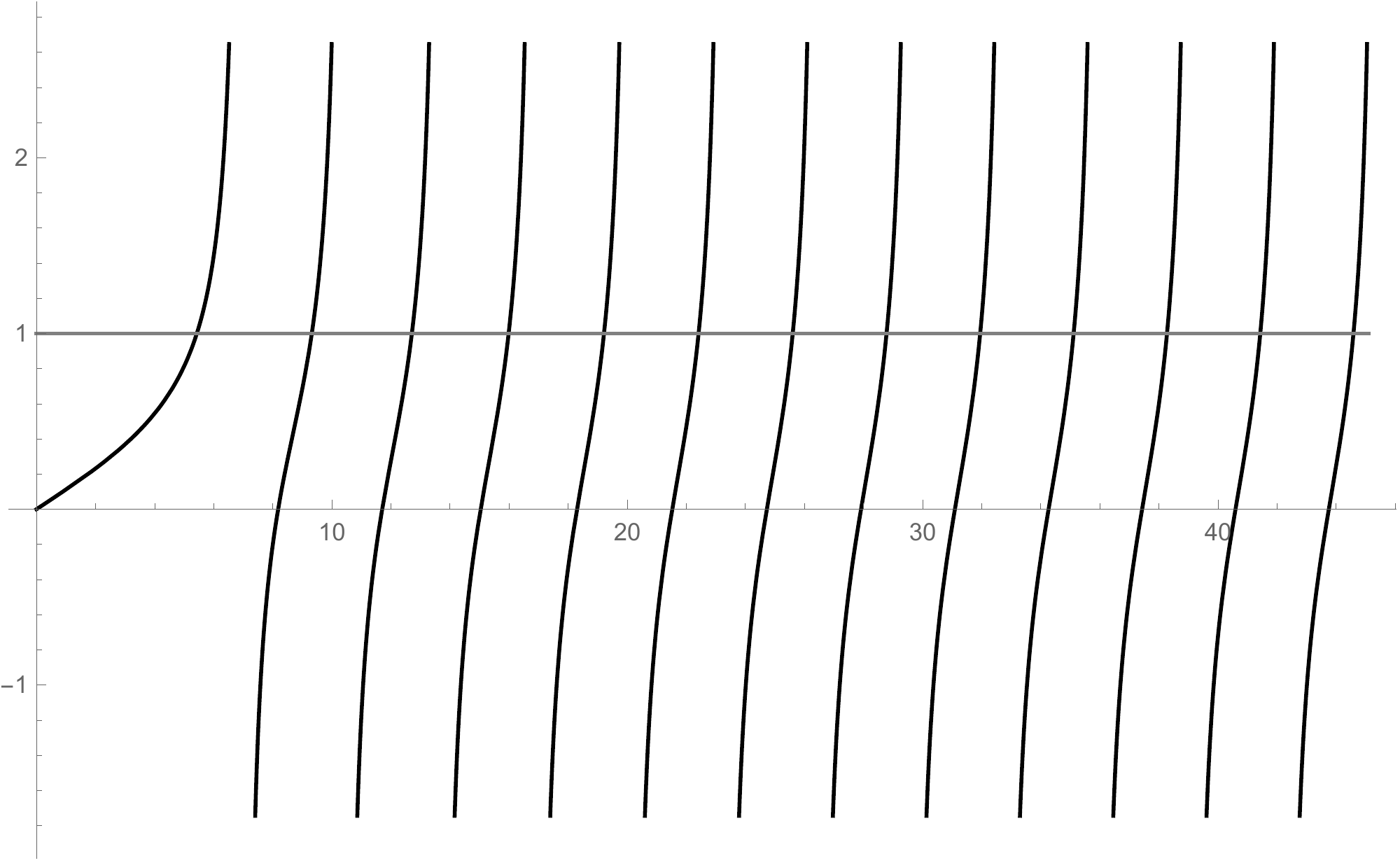}
\caption{\label{tha_label} Solutions to Eq. 
(\ref{lambda}) for $j=\frac 72,\ c=1$.}
\end{center}
 \end{figure}

Let us  check that eigenfunctions (\ref{resenje}) can satisfy
 (\ref{Key}). First relation is clearly true, the second gives
\begin{equation}
  \frac{J_{j+1}(2\lambda)}{J_j(2\lambda)}= c ={\rm const} ,  
\label{lambda}
\end{equation}
that is,  an equation for $\lambda$. This equation, as 
seen from Figure 1, has infinitely
many solutions  for every real $c$; the set of solutions is 
discrete. The other way to see this is for large values of 
$\lambda$ as, asymptotically,
\begin{equation}
   \frac{J_{j+1}(2\lambda)}{J_j(2\lambda)} \sim-\tan \left( 2\lambda -\frac{(2j+1)\pi}{4}\right) , \qquad \lambda\to\infty \,
.                 \label{tan}
\end{equation}
The eigenvalues  can be labelled by a natural number $n$, and
for large $\lambda$ they become equidistant with period
$\,\pi/ 2\,$. By a choice of   $c\,$ we can fix the value 
of one of the $\lambda$'s; the other eigenvalues are 
determined by $(\ref{lambda})$. This means that
for every $c$ we obtain a different self-adjoint extension 
 $\, T^{(c)}_{j}$, i.e. we have a one-parameter family: 
we can take  for example $\,c=1$ as a preferred choice.

Let us check orthogonality. Using the recurrence 
relations between the Bessel functions we find
\begin{eqnarray}
 && (\psi_{\lambda j m},\psi_{\lambda'j'm'}) = 
2 C^*C\delta_{jj'}\delta_{mm'}\int\limits_0^1z\, dz \left(
J_j(2\lambda z)J_j(2\lambda' z)+J_{j+1}(2\lambda z)J_{j+1}(2\lambda' z)
\right) \nonumber \\
&& \quad =
\delta_{jj'}\delta_{mm'}\, \frac{\vert C\vert^2}{\lambda^2-{\lambda'}^2}\,
\Big( \lambda'J_j(2\lambda)  J'_j(2\lambda')
+\lambda'J_{j+1}(2\lambda) J'_{j+1}(2\lambda')\nonumber
\\
&&\phantom{\quad =
\delta_{jj'}\delta_{mm'}\, \frac{\vert C\vert^2}{\lambda^2-{\lambda'}^2\ }
\Big(}
-\lambda J_j(2\lambda')  J'_j(2\lambda)
-\lambda J_{j+1}(2\lambda')  J'_{j+1}(2\lambda)\Big)\nonumber \\[4pt]
&&\quad
=-\, \frac{\delta_{jj'}\delta_{mm'}}{\lambda-\lambda'}\,
\frac{\vert C\vert^2}{J_{j+1}(2\lambda) J_{j+1}(2\lambda')}\,
\left(\frac{J_j(2\lambda)}{J_{j+1}(2\lambda)}-
 \frac{J_j(2\lambda')}{J_{j+1}(2\lambda')} 
\right)\, ,
\end{eqnarray}
where in the second line $J'_a(\zeta)$ denotes the derivative
of $J_a(\zeta)$.
The last expression is zero for $\lambda\neq \lambda'\,$
for discrete set of eigenfunctions
which satisfy (\ref{lambda}), and we 
confirm that the given basis is orthogonal.

\initiate

\section{Singularities and symmetries}

Let us verify that fuzzy de~Sitter space corresponds to an 
expanding cosmology and discuss the absence of the big bang 
singularity. The (squared) radius of the universe is 
given by
\begin{equation}
 (x^i)^2 = - \ell^2 W_iW^i\, 
\end{equation}
and its evolution can be traced by  the
expectation value   $\,\langle (x^i)^2\rangle\, $  in the
eigenstates of  time. 
Eigenvalue $\lambda\,$ of $\, W_0-W_4\,$ used
in the previous calculation is related to the time eigenvalue 
 $t$ exponentially,
\begin{equation}
t =\langle \tau\rangle =(\psi_{\lambda jm}, \tau \,\psi_{\lambda jm})
 = \ell\, \log\lambda\, .
\end{equation}
 Using  Casimir relation (\ref{Casimir}),
\begin{equation}
 -W_iW^i = {\cal W} +W_0^2-W_4^2 \, ,
\end{equation}
 and taking normalized eigenstates $\, \psi_{\lambda j m}$
\begin{equation}
(\psi_{ \lambda jm},  \psi_{\lambda jm}) 
= 2 C^*C\int\limits_0^1 dz \, z \left(J_j^2(2\lambda z)+ 
J_{j+1}^2(2\lambda z)\right) =1 \,                  \label{norm}
\end{equation}
we find
\begin{eqnarray}
\langle - W_iW^i\rangle = 
{\cal W}+\langle(W_0+W_4) (W_0-W_4) \rangle
= {\cal W}+ \lambda^2+ 2\lambda\, \langle
W_4\rangle\, .  \nonumber
\end{eqnarray}
Expectation value $\,\langle  W_4\rangle\, $ can be  estimated.
We have $\  W_4= -\frac 12
\begin{pmatrix}
 p_0\vec r\cdot \vec\sigma & i\vec L\cdot\vec\sigma +\frac{3i}{2} \\
i\vec L\cdot\vec\sigma +\frac{3i}{2}  & p_0\vec r\cdot \vec\sigma
\end{pmatrix}\, $,
 therefore
\begin{eqnarray}
 &&\hskip-1cm  (\psi_{ \lambda jm}, 
W_4\,\psi_{\lambda jm}) =
\int \frac{d^3p}{p_0}\, \Phi_{\lambda jm}^\dagger
\left(\frac{im\, \vec p\cdot\vec\sigma}{2(p_0+m)^2}\,  
-\frac{m^2\,\vec r\cdot\vec\sigma}{p_0+m}\,  -
 \frac{im\, (\vec p\cdot\nabla )(\vec p\cdot\vec\sigma)}{(p_0+m)^2}\,
\right) \Phi_{\lambda jm}        \nonumber \\
&&  \hskip-1cm 
\phantom{ \langle\lambda jm\vert\, W_4\,\vert\lambda jm\rangle} 
= -\frac i2\int\limits_0^1 dz\, (1-z^2)\left( F_{\lambda j}^*\,
\frac{dH_{\lambda j}}{dz}
+ H_{\lambda j}^*\, \frac{dF_{\lambda j}}{dz}\right)
 \nonumber \\
&&\hskip-1cm
 \phantom{ \langle\lambda jm\vert\, W_4\,\vert\lambda jm\rangle} 
=\lambda\, C^*C\int\limits_0^1 dz \, z(1-z^2) \left(J_j^2(2\lambda z)+ 
J_{j+1}^2(2\lambda z)\right)    .
\end{eqnarray}
Comparing the last integral with (\ref{norm}),
\begin{equation}
0\leq \int\limits_0^1 dz \, z(1-z^2) \left(J_j^2(2\lambda z)+ 
J_{j+1}^2(2\lambda z) \right) \leq
 \int\limits_0^1 dz \, z \left(J_j^2(2\lambda z)+ 
J_{j+1}^2(2\lambda z)\right)  
\end{equation}
we obtain that $  \ 
0\leq  (\psi_{ \lambda jm}, 
W_4\,\psi_{\lambda jm}) \leq
\frac \lambda 2 \ $,
hence
\begin{equation}
 {\cal W} +\lambda^2\leq (\psi_{ \lambda jm},  - W_iW^i\, 
\psi_{\lambda jm})  \leq  {\cal W}+ 2\lambda^2
 \, .
\end{equation}
The expectation value of the radius of the universe 
is bounded below by $\ \ell\sqrt{{\cal W}}\,$:
it does not vanish in physical states which lie in the
domain of self-adjoint  extensions  $\tau^{(c)}$, i.e.
can be expanded in the corresponding
 eigenbases.  The radius,  on the other hand, grows with 
time exponentially:  for late times we have 
$\, \sqrt{\langle - W_i W^i\rangle}\sim \lambda = 
e^{t/\ell}\,$.

Another important point is discreteness of time
that, as explained, also comes through the
self-adjointness of $\, \tau$. Though hermiticity
 is  a usual  condition in quantum mechanics, we 
rarely deal  with operators that do not
 have unique self-adjoint extensions.
This is  related to the fact that quantum mechanics is defined 
on the flat unbounded  space: one can expect boundary effects 
in curved spaces, spaces which are bounded or singular
(geodesically incomplete, or with curvature singularities).
In this context, formally symmetric hamiltonians with a one-parameter
family of self-adjoint extensions appear in various physical 
situations (and mathematical setups) in general relativity and cosmology, \cite{Wald:1980jn,Horowitz:1995gi,Andrianov:2018wdx,Gryb:2018whn}.
The interpretation of non-uniqueness of the extension varies:
 from understanding that it is a further quantization ambiguity
\cite{Andrianov:2018wdx}, 
to that  it renders a definition of spacetimes that are singular for 
 `quantum probes' (as in some cases, classically singular
spacetimes can appear completely regular for quantized particles),
 \cite{Horowitz:1995gi}.
Wald relates the necessity to choose one of the  extensions
 with the fact that the initial-value problem is classically
 ill-defined at naked singularity,
and regards the possibility of constructing a self-adjoint extension
as a resolution to the singularity problem, \cite{Wald:1980jn}.

The last point of view is in some sense close to our example, 
though we are extending  time and not the hamiltonian.
Discreteness of time becomes relevant in the
`deep quantum region' $\,\lambda\to\, 0\,$, i.e. $t\to -\infty\,$,
near the classical boundary through which the steady-state model
can be  extended to the complete de~Sitter space.
For values away from the Planck scale
time is almost continuous: the difference 
between its consecutive eigenvalues is macroscopically negligible,
\begin{equation}
t_{n+1}-t_n \approx \ell\, \log\, (1+\frac 1n\,)  \, .
\end{equation}



Discreteness  obtained by requiring self-adjointness 
in known in other cases of quantum spaces. One example is
the $q$-deformed Heisenberg algebra,
\begin{equation}
 [p,x] = -i +(q-1)xp \, 
\end{equation}
 and its unitary representations, \cite{Schwenk:1992sq,Hebecker:1993eb}. 
The analysis shows that coordinate $x$ is not self-adjoint,
but the self-adjoint extensions exist; both $x$ and $p$ have discrete spectra. 
Another interesting  case is the minimal-length  Heisenberg algebra, 
\begin{equation}
 [p,x]=-i-i\beta p^2 ,
\end{equation}
which is in \cite{Kempf:1994su}  represented in the
 Schr\"odinger representation. Again it is
 found that $x$ has a one-parameter family of self-adjoint
extensions which puts its spectrum on  lattice. 

The $q$-deformed Heisenberg algebra (Manin plane)
has, as symmetry,  the quantum group $SU_q(2)$, so
it is natural to ask whether in our model 
symmetry gets deformed as well.
As shown in \cite{Buric:2015wta}, our choice of 
 frame in fact breaks the $SO(1,4)$ invariance, and 
 {\it a priori} symmetries of  fuzzy de~Sitter space
are rotations and time translations,  $SO(3)\times U(1) $.
Here $U(1)$  denotes the dilatation subgroup,
 $\,U(1) =\{  e^{i\alpha M_{04}}\vert \alpha\in\mathbb{R}\}\,$,
the dilatation generator plays the role of the hamiltonian, 
$\,H=M_{04}\,$: it evolves the eigenstates of time,  (\ref{dilatation}).

If we keep the standard notion that symmetry is
defined  by  group of transformations, 
the choice of a self-adjoint extension  $\tau^{(c)}\,$  
is spontaneous symmetry breaking. This can 
be seen easily: the elements of $U(1)$   do not preserve 
the space of physical states defined by  (\ref{Key}) for 
arbitrary values of parameter $\alpha$. However, there is a
subgroup of dilatations,  $U^{(c)}(1)$,  
determined by the allowed values of $\alpha$ which preserve
condition (\ref{Key}): it is  represented 
nonlinearly. For large eigenvalues, (\ref{Key}) 
becomes periodic and   $\lambda$  equally spaced: 
subgroup $U^{(c)}(1)$ becomes in this limit (in this region 
of physical parameters),  the additive group of integers.
In the  continuum approximation $\,\ell\to 0\,$ which is valid
on the macroscopic scale,  the full symmetry is recovered.
Another  view is that, in the
quantum regime,  classical symmetries get deformed, 
\cite{Fichtmuller:1995dt}: whether the corresponding
transformations in our case have  the structure of
a quantum group is to be studied.
In any case, what we find is that classical symmetries
get broken or deformed on the Planck scale,
 due to the quantum structure of spacetime. 
To obtain other effects in cosmology which our model
predicts we should introduce matter, for
example  scalar field. This is in principle a well defined problem 
in noncommutative geometry and  we plan to address it in our future work.

\vskip0.5cm
\begin{large}
{\bf Acknowledgement}
\end{large}
We thank Igor Salom and Ilija Buri\' c for various discussions
on representation theory. The work was supported by the Serbian 
Ministry of Education, Science and Technological Development Grant ON171031, 
and by the COST action MP 1405 ``Quantum structure of spacetime''.

\vskip1cm

\begin{large}
{\bf Appendix 1}
\end{large}

In this appendix we solve the radial equations (\ref{Eigen}). 
In the signature which we use
\begin{eqnarray}
 && \vec p =(p_i),\quad \vec L = (L_i), \quad \vec\sigma = (\sigma_i),\quad
\vec r= (x^i) = i\,\frac{\p}{\p p_i}\, , \nonumber 
\\
&&\vec p\cdot \vec\sigma = -p_i\sigma^i  ,\quad \sigma_i\sigma_j=-\eta_{ij}-\epsilon_{ijk}\sigma^k \quad ,
(\vec r\cdot \vec \sigma)(\vec p\cdot\vec\sigma) = i\Big(
 3+p\, \frac{\p}{\p p}+\vec L\cdot \vec\sigma\Big)   .
\nonumber
\end{eqnarray}
The eigenvalue equation (\ref{Eigen}) is
\begin{equation}
 \Big( \frac{1}{2m}\,\rho\,(\vec p\cdot\vec\sigma) - \frac 12
(p_0+m)\,(\vec r\cdot \vec \sigma) -\frac{1}{2m}
(\vec p\cdot \vec r)(\vec p\cdot \vec\sigma) \Big) \Phi =\lambda\Phi  \,      .        \label{eigen}
\end{equation}
We use the Ansatz  which separates 
angular and radial variables,
\begin{equation}
\Phi_{\lambda jm}(\vec p) = \frac{f(p)}{p}\,
\phi_{jm}(\theta,\varphi)+\frac{h(p)}{p}\,
\chi_{jm}(\theta,\varphi)  \, ,
\label{ansatz}
\end{equation}
with  $\, p^2 =-p_i p^i= p_0^2 -m^2\, $.
The $\, \phi_{jm} $ and  $ \chi_{jm} $ are the
 spinor eigenfunctions of  $\,M_{ij}M^{ij}$ and
$\,M_{12}$; they are orthonormal and satisfy
\begin{equation}
 \begin{array}{ll}
    \phi_{jm} =\dfrac{\vec p\cdot\vec\sigma}{p} \,  \chi_{jm}\, ,\quad\ 
&  (\vec L\cdot\vec\sigma)\, \phi_{jm} = (j-\frac 12)\,\phi_{jm}\, ,
  \\[8pt]
\chi_{jm} =\dfrac{\vec p\cdot\vec\sigma}{p} \,  \phi_{jm} \, ,
\quad
&
(\vec L\cdot\vec\sigma)\, \chi_{jm} = -(j+\frac 32)\,\chi_{jm}  \,  .
\end{array}
\end{equation}
Introducing (\ref{ansatz}) we  obtain radial equations
\begin{align}
& (p_0+1)\, \frac{df}{dp_0}+ i\rho f-\frac{ j+\frac 12 }{p_0-1}\,
  f =2i\lambda\, \frac hp \, , \label{1}     \\[4pt]
&  (p_0+1)\, \frac{dh}{dp_0}+ i\rho h+\frac{ j+\frac 12 }{p_0-1}\,
  h =2i\lambda\, \frac fp   
\, .           \label{2}
\end{align}
In order to simplify them we rescale momentum 
to be dimensionless, $p\to mp$, \  $p_0\to mp_0$, $p\in (0,\infty)$, 
$p_0\in (1,\infty)$. Equations  decouple when we
 introduce new functions  $\,F$, $H\,$ by
\begin{eqnarray}
 f=(p_0+1)^{-i\rho-\frac{2j+1}{4}}\, (p_0-1)^{\frac{2j+1}{4}} \, F ,\quad\ 
 h=(p_0+1)^{-i\rho+\frac{2j+1}{4}}\, (p_0-1)^{-\frac{2j+1}{4}} \, H  .
                                              \label{newfunctions}
\end{eqnarray}
We then obtain
\begin{align}
& (p_0^2-1) \,\frac{d^2 F}{dp_0^2} +2(p_0+j)\,\frac{d F}{dp_0}
 + \frac{4\lambda^2}{(p_0+1)^2}\, F =0      \, ,                 \label{A}\\[6pt]
&(p_0^2-1) \,\frac{d^2 H}{dp_0^2} +2(p_0-j-1)\,\frac{d H}{dp_0}
 + \frac{4\lambda^2}{(p_0+1)^2}\, H =0    \,,
 \,                                                                   \label{B}
\end{align}
and  additional relations
\begin{equation}
 \frac{dF}{dp_0} = 2i\lambda(p_0+1)^{j-1}(p_0-1)^{-j-1}\, H, \quad \ 
 \frac{dH}{dp_0} = 2i\lambda(p_0+1)^{-j-2}(p_0-1)^{j}\, F  .     \label{FiH}
\end{equation}

Equations (\ref{A}-\ref{B}) reduce to the Bessel equation
\begin{equation}
 \zeta^2\, \frac{d^2Y}{d\zeta^2} +\zeta\, \, \frac{d Y}{d\zeta}
 +(\zeta^2-a^2)Y=0                       \label{bessel}
\end{equation}
by compactification of the independent variable. 
Introducing  $z$ as
\begin{equation}
 z=\sqrt{\frac{p_0-1}{p_0+1}}\ 
\end{equation}
 both equations reduce to (\ref{bessel})
for $\,\zeta = 2\lambda z\in (0, 2\lambda)$.  In equation (\ref{A}),  
$a=j\,$;  in (\ref{B}),  $a=-j-1$. 

Linearly independent solutions to the Bessel equation
are the Bessel functions $\,J_a(\zeta)$, $\,J_{-a}(\zeta)\,$ or 
$J_a(\zeta)$, $\,Y_{a}(\zeta)$  ($a$ is half-integer, so
$\,J_{-j-1}(\zeta) = (-1)^{j-\frac 12}Y_{j+1}(\zeta)$).
Therefore, $\, F\sim J_j,\, J_{-j}$ and $\, H\sim J_{j+1},\, J_{-j-1}$.
 Taking into account additional relations
(\ref{FiH}) which are satisfied through recurrence relation
\begin{equation}
 \frac 1\zeta\, \frac{d}{d\zeta}\zeta^a J_a(\zeta)) =
 \zeta^{a-1} J_{a-1}(\zeta) \, ,
\end{equation}
 we obtain two linearly independent solutions:
\begin{eqnarray}
 & F_{\lambda j} =Cz^{-j} J_j(2\lambda z)\,,\ \ \ 
& H_{\lambda j} =i Cz^{j+1} J_{j+1}(2\lambda z)\label{sol}\, , \\[4pt]
& \ \,\tilde F_{\lambda j} =\tilde Cz^{-j} J_{-j}(2\lambda z)\, ,\quad 
& \tilde H_{\lambda j} =-i \tilde C z^{j+1} J_{-j-1}(2\lambda z) \, 
                                            .                  \label{sol'}
\end{eqnarray}
As the Bessel functions around $\,\zeta =0\,$ behave as
\begin{equation}
 J_a(\zeta)\sim \frac{1}{\Gamma(a+1)}\, \left(\frac{\zeta}{2}\right)^a \, 
\end{equation}
the second solution diverges, 
$\ \tilde\psi_{\lambda jm}\sim \zeta^{-j-\frac 32}\, $,
so we have one regular solution,
\begin{equation}
 f_{\lambda j} = C\left(\frac{2}{1-z^2}\right)^{-i\rho} \sqrt{z}\, 
J_j(2\lambda z),\quad \ 
h_{\lambda j} = iC\left(\frac{2}{1-z^2}\right)^{-i\rho} \sqrt{z}\, 
J_{j+1}(2\lambda z) .
\end{equation}
It exists for every real  $\lambda$. But 
 $\, J_a(-\zeta) = (-1)^{a}J_a(\zeta) $, 
so the spectrum  can be restricted to the 
positive real axis,  $\lambda>0$.

The scalar product of  two eigenfunctions is given by
\begin{eqnarray}
&& (\psi_{ \lambda jm},\psi_{\lambda' j'm'}) =2\delta_{jj'}\delta_{mm'}
\int\limits_0^1 dz\,(f^* f'+h^*h')= 2\delta_{jj'}\delta_{mm'}
\int\limits_0^1 dz\,(z^{2j+1}F^* F'+z^{-2j-1}H^*H') 
\nonumber
\\
&&\phantom{ \psi_{\lambda jm} }
 =2\delta_{jj'}\delta_{mm'}C^*C'
\int\limits_0^1z dz\, \left( J_j(2\lambda z)J_j(2\lambda'z)+
 J_{j+1}(2\lambda z)J_{j+1}(2\lambda'z) \right) \, .       \label{proizvod}
\end{eqnarray}
It is nonzero for $\lambda\neq \lambda'$, and finite for each
$\lambda$, which as we discuss in the text, is a problem.
Singular solutions do not have the right normalization 
to be eigenfunctions of the
continuous spectrum:  similarly  to (\ref{proizvod}), we have
\begin{equation}
 (\tilde\psi_{\lambda jm},\tilde\psi_{ \lambda' j'm'}) =2\delta_{jj'}\delta_{mm'}
\tilde C^*\tilde C'
\int\limits_0^1 dz\,\left( J_{-j}(2\lambda z)J_{-j'}(2\lambda'z)+
 J_{-j-1}(2\lambda z)J_{-j'-1}(2\lambda'z) \right) \, .         \nonumber
\end{equation}
This integral is divergent in the lower limit, but the divergence
depends on  $j$ and not on the difference
 $\,\lambda-\lambda'\, $ i.e. it does not have the 
required form $\,\delta(\lambda -\lambda')\,$.

\vskip1cm
\begin{large}
{\bf Appendix 2}
\end{large}

We start with  the deficiency indices of $T_{j}$. 
To determine them we need to solve equations  
\begin{equation}
 T_{j}\Phi =\pm i\Phi \, .                \label{deficiency}
\end{equation}
This is in fact not difficult: solutions to these 
equations are the same as solutions to (\ref{eigen})
for $\lambda =\pm i$: the Bessel functions of  imaginary 
argument i.e. the modified Bessel functions $I_a(\zeta)$ and
 $K_a(\zeta)$,
\begin{equation}
 I_a (\zeta) = i^{-a} J_a (i\zeta), \qquad 
K_a(\zeta)= \frac \pi 2\, i^{a+1} \left( J_a(i\zeta) +i Y_a(i\zeta)\right) \, .
\end{equation}
As before, $a=\pm j, \pm (j+1)$.
The modified Bessel functions have similar behavior around
zero as the Bessel functions: $K_a(\zeta)$ is divergent and the 
corresponding solution has infinite norm. This implies that equation
$ \, T_{j}\Phi = i\Phi \, $ has just one regular solution,
\begin{equation}
 F_+= C z^{-j} I_j(2z)\,, \qquad  H_+=- C z^{j+1} I_{j+1}(2z) \, .
\end{equation}
Similarly there is one regular solution  $\,(F_-,H_-)$
to equation  $ \, T_{j}\Phi =- i\Phi \, $. This means
that  deficiency indices of $\,T_{j}$ are $\,(n_+,n_-)=(1,1)$, 
hence $T_{j}$ is not a  self-adjoint operator but  can be
extended to one.

Next, let us briefly  recall  the procedure of constructing 
self-adjoint extensions of formally symmetric operators.
We use notation  of \cite{pim}, where also proof of the  
main technical result which we use can be found.
We can write equation (\ref{bound}) abstractly as
\begin{equation}
 (\Phi, T_j\Phi') = (T_j\Phi, \Phi') +  B(\Phi,\Phi')
=(T_j^\dagger \Phi,\Phi') \, ,
\end{equation}
where the boundary term $B(\Phi,\Phi') $  is a bilinear form,
which in our case reads
\begin{equation}
 B(\Phi,\Phi')=(F^*H'+H^*F')\lvert_0^1 \, .          \label{bound.term}
\end{equation}
Apparently,  the domain of $\,T_{j}\,$ is given by
all normalizable functions $\,\Phi\,$, $\Phi'\,$ 
that satisfy $\,B(\Phi,\Phi')=0$, or  in our case
$\,F(0)$=$H(0)$=0, $\ F(1)$=$H(1)$=0. 
Then ${\cal D}(T_{j}^\dagger) ={\cal H}\,$ and obviously
the two domains are not equal,  ${\cal D}(T_{j})\subset \cal{H}$.
To achieve self-adjointness, one should relax the condition
which determines ${\cal D}(T_{j}) $ and restrict 
${\cal D}(T_{j}^\dagger)$. 
This is done effectively by finding $\,n_+\,$ linearly
 independent functions  $\, \Phi_k\,$ (more precisely, $n_+\,$ linearly
 independent vectors corresponding to their boundary values),
 in our case one, $\Phi_1$, that satisfy
\begin{equation}
 B(\Phi_k,\Phi_l) =0.        \qquad \forall\, k,l   \,.        \label{NN}
\end{equation}
The domain of a self-adjoint extension of $\,{T}_{j}$ is 
then defined as a set of functions $\Phi$,
\begin{equation}
 {\cal D}({T}_{j}) = {\cal D}({ T}_{j}^\dagger) 
= \{ \Phi\, \vert\,  B(\Phi, \Phi_k)=0, \, \forall\, k\} \, .
\end{equation}

In principle, boundary term (\ref{bound}) is a combination
of values at both boundary points but often the
 constraints can be imposed separately. It is possible 
to do it in our case as wel:  we can choose 
$F(0)=H(0)=0 $,  in accordance
with behavior of the eigenfunctions  of $\,\tau$
which constitute a basis.   If, at the other boundary, we denote 
the values of $\,\Phi_1$ as
\begin{equation}
 F_1(1)=\sigma\, e^{i\beta}, \ \  H_1(1) = i\sigma' \, e^{i\beta'},
\end{equation}
we find $i\beta = i\beta' + n\pi$. 
Constants $\beta$, $\sigma$ and $\sigma'$ are real numbers, 
so the domain of the self-adjoint extension $\,T_j^{(c)}\,$ is a set of 
functions that satisfy
\begin{equation}
F(0)=H(0)=0, \qquad 
 H(1) = \pm \, i\, \frac{\sigma}{\sigma'}\, F(1) = ic F(1)\, ,
\quad c\in\mathbb{R}\, .
  \label{key}
\end{equation}


\begin{thebibliography}{99}



\bibitem{book}
  J.~Madore,
  ``An Introduction To Noncommutative Differential Geometry And Its Physical
  Applications,''
  Lond.\ Math.\ Soc.\ Lect.\ Note Ser.\  {\bf 257} (2000).

\bibitem{Buric:2006di}
  M.~Buric, T.~Grammatikopoulos, J.~Madore and G.~Zoupanos,
  JHEP {\bf 0604} (2006) 054
  [hep-th/0603044].


\bibitem{fs}
  J.~Madore,
  Class.\ Quant.\ Grav.\  {\bf 9} (1992) 69,\ \
J. Hoppe,  ``Quantum theory of a massless
relativistic surface and a two-dimensional bound state problem'',
Ph.D. Thesis, MIT, 1982.


\bibitem{Balachandran:2005ew}
  A.~P.~Balachandran, S.~Kurkcuoglu and S.~Vaidya,
  Singapore, Singapore: World Scientific (2007) 191 p.
  [hep-th/0511114].

\bibitem{Buric:2015wta}
  M.~Buric and J.~Madore,
  Eur.\ Phys.\ J.\ C {\bf 75} (2015) no.10,  502
  [arXiv:1508.06058 [hep-th]].

\bibitem{Buric:2017yes}
  M.~Buric, D.~Latas and L.~Nenadovic,
  Eur.\ Phys.\ J.\ C {\bf 78} (2018) no.11,  953
  [arXiv:1709.05158 [hep-th]].

\bibitem{Hawking}
 S. W. Hawking and G. F. R. Ellis,
``Large Scale Structure of Space-Time'',
 Cambridge University Press (1975)

\bibitem{Buric:2006nr}
  M.~Buric, D.~Latas, V.~Radovanovic and J.~Trampetic,
  Phys.\ Rev.\ D {\bf 75} (2007) 097701
  doi:10.1103/PhysRevD.75.097701
  [hep-ph/0611299].




\bibitem{R}
J.~Dixmier,
Bull. Soc. Math. France 89 (1961) 9.



\bibitem{Moylan}
  P.~Moylan,
  J.\ Math.\ Phys.\  {\bf 24} (1983) 2706,
  P.~Moylan,
  J.\ Math.\ Phys.\  {\bf 26} (1985) 29.


\bibitem{Bargmann:1948ck}
  V.~Bargmann and E.~P.~Wigner,
  Proc.\ Nat.\ Acad.\ Sci.\  {\bf 34} (1948) 211.
%
%

\bibitem{pim}
V. Hutson, J. Pym, M. J. Cloud,
``Applications of Functional Analysis and Operator Theory'',
 Elsevier Science (2005)



\bibitem{dunford}
N. Dunford and J. T. Schwartz,   ``Linear Operators'', 
J. Wiley \& Sons (1957).
 
%

\bibitem{Wald:1980jn}
  R.~M.~Wald,
  J.\ Math.\ Phys.\  {\bf 21} (1980) 2802.
  doi:10.1063/1.524403

\bibitem{Horowitz:1995gi}
  G.~T.~Horowitz and D.~Marolf,
  Phys.\ Rev.\ D {\bf 52} (1995) 5670
  doi:10.1103/PhysRevD.52.5670
  [gr-qc/9504028].


\bibitem{Andrianov:2018wdx}
  A.~A.~Andrianov, C.~Lan, O.~O.~Novikov and Y.~F.~Wang,
  Eur.\ Phys.\ J.\ C {\bf 78} (2018) no.9,  786
  doi:10.1140/epjc/s10052-018-6255-5
  [arXiv:1802.06720 [hep-th]].



\bibitem{Gryb:2018whn}
  S.~Gryb and K.~P.~Y.~Th\' ebault,
  Class.\ Quant.\ Grav.\  {\bf 36} (2019) no.3,  035009
  doi:10.1088/1361-6382/aaf823
  [arXiv:1801.05789 [gr-qc]].


%
%
%


\bibitem{Schwenk:1992sq}
  J.~Schwenk and J.~Wess,
  Phys.\ Lett.\ B {\bf 291} (1992) 273.

\bibitem{Hebecker:1993eb}
  A.~Hebecker, S.~Schreckenberg, J.~Schwenk, W.~Weich and J.~Wess,
  Z.\ Phys.\ C {\bf 64} (1994) 355.

\bibitem{Kempf:1994su}
  A.~Kempf, G.~Mangano and R.~B.~Mann,
  Phys.\ Rev.\ D {\bf 52} (1995) 1108
  [hep-th/9412167].


\bibitem{Fichtmuller:1995dt}
  M.~Fichtmuller, A.~Lorek and J.~Wess,
  Z.\ Phys.\ C {\bf 71} (1996) 533
  [hep-th/9511106].


%
\end{thebibliography}
\end{document}